\documentclass[12pt,twocolumn,fleqn]{article}
\usepackage[dvips]{graphicx}
\usepackage{epsfig}

\newcommand{\AmS}{{\protect\the\textfont2
  A\kern-.1667em\lower.5ex\hbox{M}\kern-.125emS}}

\title{Multifragmentation in central collisions \\ 
at intermediate energies: a fast process?}

\author 
{A. Van Lauwe $^a$, D. Durand~$^a$,
G.~Auger~$^b$,
N.~Bellaize~$^a$,\\
B.~Bouriquet~$^b$,
A.M.~Buta~$^a$,
J.L.~Charvet~$^d$,\\
A.~Chbihi~$^b$,
J.~Colin~$^a$,
D.~Cussol~$^a$,
R.~Dayras~$^d$,\\
A.~Demeyer$^f$,
J.D. Frankland~$^b$,
E.~Galichet~$^{c,g}$,
D.~Guinet~$^f$,\\
B.~Guiot~$^b$,
S.~Hudan~$^b$,
G.~Lanzalone$^{c}$,
P.~Lautesse~$^f$,\\
F.~Lavaud~$^d$,
J.F.~Lecolley~$^a$,
R.~Legrain~$^{d,\dagger}$,
N.~Le Neindre~$^b$,\\
O.~Lopez~$^a$,
L.~Manduci$^a$,
J.~Marie$^a$,
L.~Nalpas~$^d$,\\
J.~Normand~$^a$,
M. P\^arlog~$^{c,h}$,
P.~Paw{\l}owski$^2$,
E.~Plagnol~$^c$,\\
M.F.~Rivet~$^c$,
E.~Rosato~$^e$,
R.~Roy$^i$,
J.C.~Steckmeyer~$^a$,\\
G.~T\u{a}b\u{a}caru~$^{c,h}$,
B.~Tamain~$^a$,
E.~Vient~$^a$,
M.~Vigilante~$^e$,\\
C.~Volant~$^d$
J.P.~Wieleczko~$^b$
\\
\\
\bf{(INDRA collaboration)} \\
\\
$^a$ Laboratoire de Physique Corpusculaire de Caen, France \\
$^b$ Grand Acc\'el\'erateur National d'Ions Lourds, Caen, France\\
$^c$ Institut de Physique Nucl\'eaire, Orsay, France \\
$^d$ DAPNIA/SPhN, CEA/Saclay, France\\
$^e$ Dipartimento di Scienze Fisiche e Sezione INFN,
Universit\`a di Napoli, Italy\\
$^f$ Institut de Physique Nucl\'eaire, Lyon, France\\
$^g$ Conservatoire National des Arts et M\'etiers, Paris, France\\
$^h$ National Inst. for Physics and Nucl. Engineering, Bucharest, Romania\\
$^i$ Laboratoire de Physique Nucléaire, Université Laval, Québec,Canada.\\
\\}

\begin{document}
% typeset front matter

\maketitle

\begin{abstract}
The kinematical characteristics of fragments and light particles 
observed in central highly fragmented nuclear
collisions at intermediate energies are compared with the results 
of a model assuming that 
the initial momentum distribution of the
nucleons inside the two partners of the reaction has no time to 
relax before the disassembly of the system.
The rather good agreement between the results of the model and the experimental data suggests that
multifragmentation is a fast process with a strong memory of the
entrance channel.  
\end{abstract}
\section{Introduction}
In central nuclear collisions at intermediate energies, multifragmentation, defined as the emission in a short time scale 
of several species 
of atomic number larger than two, is an important process as compared to other decay mechanisms 
such as the formation of heavy residues or fission (see for instance \cite{intro}). Such reactions are thus believed to be the ideal tool 
to study the transition from a liquid-like state (nuclei at normal density) towards a gas-like state associated 
with the vaporisation of the system \cite{vapo}. In such a picture, nuclear multifragmentation is 
associated with the excursion of the excited
system at low densities inside the coexistence region of the nuclear phase diagram \cite{chocho}. There,  
a volume instability, the so-called spinodale
decomposition, leads the system to fragmentation. Such a scenario is supported by microscopic 
transport calculations
based on the nuclear Boltzmann equation and its stochastic extensions \cite{bnv}. These models   
predict a compression-expansion 
phase driving the system at low density.
 
Recent experimental analyses based on nuclear calorimetry have claimed evidence for a liquid-gas phase transition 
through the study of various signals \cite{bebor,dagostino,bougault}. Some of these analyses make extensive use 
of the thermal multifragmentation statistical models \cite{bondorf,gross}
to prove the existence of thermal equilibrium \cite{dagostino,bougault}. In such a framework, 
charge distributions and fragment multiplicities 
are fitted with help of several parameters characterising the thermalised source. 
More involved studies are based on 'back-tracing' 
techniques and they allow to obtain not only the mean values but also 
the distributions of the parameters of the source \cite{lopez}. 
It is worth noting that, in all cases,
not all data are fitted. Indeed, it is very often assumed that a strong pre-equilibrium 
phase is present and this is why kinematical cuts are applied to the data before comparison 
with the predictions of the models.    
In such thermal models, 
the only source of fluctuations regarding the kinetic
energy distributions of the emitted species is the thermal motion inside the thermalised source 
associated with a possible (usually assumed self-similar) collective motion \cite{lavaud}. 
Recently, deformation has been introduced 
to account for the observed anisotropies in the data \cite{boubour}.
  
Our aim in the present work 
is to explore the possibilities of a scenario in which  
the main hypothesis underlying the statistical models is abandoned. 
This means that we wish to
compare the data (with as less kinematical cuts as possible) with a model taking into account explicitely the entrance
channel characteristics of the reaction as well as the initial correlations 
of the nucleons in momentum space.  
This is, of course, not the first attempt of this kind. For instance, molecular
dynamics models have been used to understand fragmentation in such 
a scenario \cite{nebauer}. 
Indeed, all molecular dynamics approaches, whether they are 'classical' or 'quantal', conclude that 
fragmentation is a fast process for which there is no time for the system to build compression or reach 
equilibrium. Here, we develop, in this spirit, a phenomenological and transparent approach allowing 
a detailed comparison with the data.     
Since the statistical models are able to correctly reproduce the 'static' properties 
of nuclear fragmentation, we focus our study  
on the 'kinematical' properties for both fragments and light particles 
obtained in central collisions in the Fermi energy range. The data used for the
study have been collected by the INDRA collaboration near the GANIL 
facility \cite{indralib}. We consider central collisions for Xe+Sn at 32 and 50 Mev/u \cite{xesnpapers}.
This contribution is organised as follows: the model (hereafter called the ELIE event generator) is described 
in section \ref{des} and a detailed comparison with experimental data is developped in section \ref{dev}. Conclusions
and perspectives are drawn at the end of the paper. 
\section{The ELIE event generator}
\label{des}
The main idea of the model is to consider the extreme hypothesis for which 
the initial momentum distribution of the nucleons has no time to relax before the fragmentation 
of the system and for which a (small) fraction of the 
nucleons suffers hard nucleon-nucleon collisions leading them to escape the system as free particles. For the 
(large) remaining part of the system, only a rearrangement of the nucleons in the final state inside the various emitted products is
considered.   
\begin{figure}[!htb]
\begin{center}
\epsfig{file=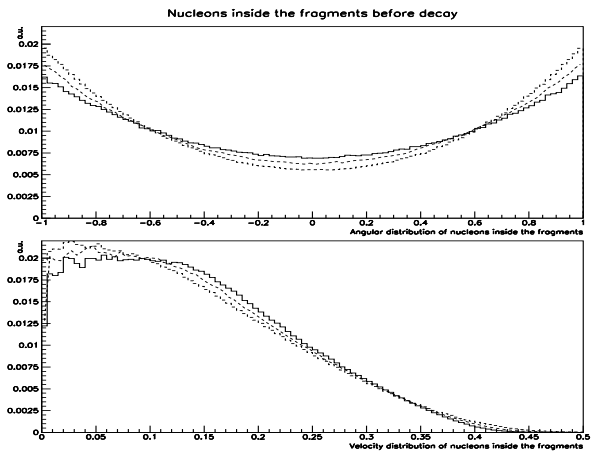,width=80mm,height=60mm}
\caption{\it  Center-of-mass angular (up) and velocity (down) distributions of nucleons inside the fragments 
before decay for Xe+Sn at 32 MeV/u (solid histograms), Au+Au at 40 MeV/u (dashed line) and Xe+Sn at 50 MeV/u (dashed histograms).} 
\label{fig0}
\end{center}
\end{figure} 
As mentioned in the introduction, we are essentially interested in the kinematical properties of multifragmentation. Therefore, the
definition of the partition (that is the 'chemical' composition in terms 
of complex particles and fragments present in the final state of the reaction) 
is not determined from {\it ab initio} calculations but we rather
use the information given by the experimental data. In other words, the mean multiplicity of each species 
is put 'by hand'.  
However, since it is assumed that the fragments are excited, the total multiplicity of detected 
particles is the sum of two
components. The first one is associated with the 'fast' production during the reaction time and the other corresponds to 
longer time scales and is
associated with the evaporation of particles by the fragments. An iteration on the magnitude of these two components is performed until 
a reasonable
agreement with the data is reached. As far as fragments are concerned, only the mean and the second moment of the 
multiplicity distribution are needed. Indeed, it appears that the charge distribution can be reproduced rather accurately by randomly 
choosing sequentially the size of the fragments (see Fig.\ref{fig3}). Once the partition is fixed, the energetics of the
process is considered. The energy balance writes:
\begin{equation}
\label{equ1}
E_{0}=E_{ko}+E_{pot}+Q+E_{K}+E^{*}
\end {equation}
where $E_{0}$ is the available centre-of-mass energy, $E_{ko}$ is the energy corresponding to 
the few knocked-out nucleons, $Q$ is the mass energy balance between the entrance channel and the considered 
partition, $E^{*}$ is the
excitation in the fragments while $E_{K}$ is the total kinetic energy carried away by 
the fragments and the light particles. Last, $E_{pot}$ is the potential energy associated with the long 
range Coulomb force acting
between the charged species of the partition. As far as light complex particles and fragments 
are considered, their composition is obtained by randomly chosing nucleons in the two initial Fermi momentum spheres 
separated by the relative momentum corresponding to the beam energy. The kinetic energy of each species 
is thus obtained by summing the momenta
of all nucleons belonging to the considered complex product. 
The total kinetic energy obtained by this procedure is such that there is
little chance to fullfill energy conservation. Thus, 
a Metropolis procedure is developped. An exchange of a nucleon 
(a Metropolis move) between two species chosen randomly is performed and the kinetic energies recalculated. 
The procedure is iterated until the value $E_{K}^{(i)}$ at step $i$ 
matches the value of $E_{K}$ needed to fulfill Eq.\ref{equ1} with a given accuracy of typically less than a percent. 
Then, the partition is ready for 
propagation in order to 'burn' the excitation energy of the fragments and to transform the potential Coulomb energy into kinetic energy.
The initial momenta are given by the procedure described above. The initial positions are obtained by randomly placing the species 
inside a volume sufficiently large so that nuclei and particles do not overlap: this gives the value of $E_{pot}$. The volume
in question is not a critical parameter since the key quantity to be compared with the data is not the separate values of 
$E_{pot}$ and $E_{K}$ but rather their
sum (which is constrained by Eq.\ref{equ1}) (see for instance \cite{adn} for a detailed discussion on this point). 
Then, the partition is propagated in a similar way as in the
SIMON event generator \cite{simon}. In particular, the decay in flight of excited species 
is considered in order to preserve space-time correlations. The generated event is then filtered
through the INDRA software filter and stored for analysis. 
\begin{figure}[!htb]
\begin{center}
\epsfig{file=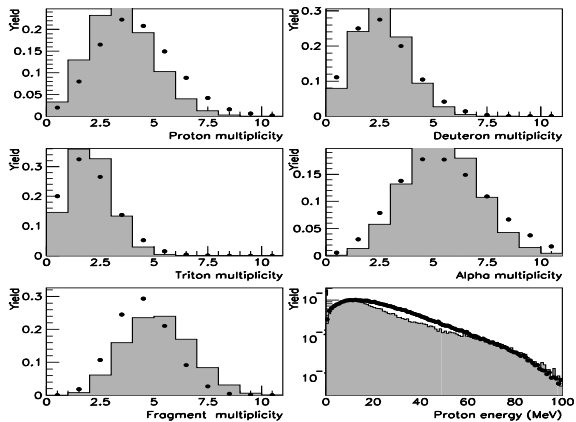,width=80mm,height=60mm}
\caption{\it  Light charged particle and fragment multiplicities 
and center-of-mass kinetic energy distributions of protons emitted in central 
Xe+Sn collisions at 50 MeV/u (see section \ref{dev}). Black points: experimental data. Filled histograms: results of the ELIE event generator.
Data have been normalized to the number of events as explained in the text.} 
\label{fig0bis}
\end{center}
\end{figure}
Apart from the fact that the partition is to a large extent put 'by hand', there remains essentially two parameters: The amount of fast
knocked-out nucleons and the mean excitation energy in each fragment. These two parameters are chosen in order to get 
the best agreement with
the experimental data (see next section). Let us discuss first the question of the excitation energy, $E^*$. 
Fig.\ref{fig0} shows the kinematical
characteristics of the nucleons inside the fragments. The angular distributions are more and more 
forward-backward peaked as the beam energy increases meaning that the fragments in this approach are more and more 
deformed and thus more and more excited. Since, in the model, 
the short range interaction among the particles is only taken into account by means of the Q-values, nucleons do not 
interact individually
and therefore the velocity distributions shown in Fig.\ref{fig0} are not physical. They however show little evolution and 
thus, it is expected that the excitation energy resulting from the deformation increases slowly with the beam 
energy.
\begin{figure}[!htb]
\begin{center}
\epsfig{file=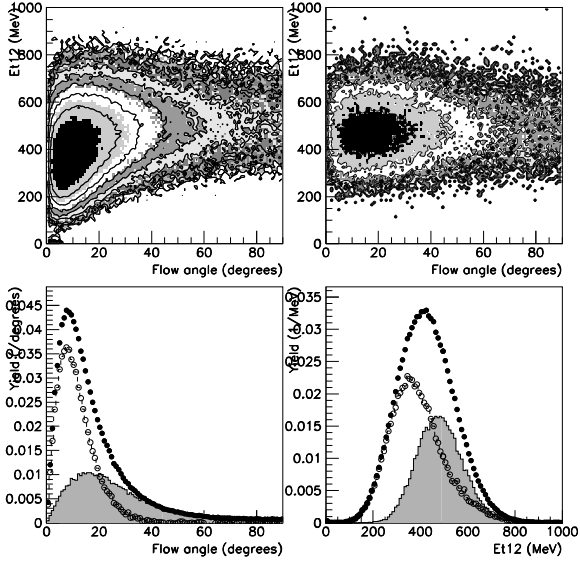,width=80mm,height=80mm}
\caption{\it  Up,left: experimental correlation between $\theta_{flow}$ and $E_{trans12}$ for Xe+Sn collisions at 
50 MeV/u. Up, right: same but for the simulation. Down, left: distribution of $\theta_{flow}$. Black points are experimental data,
the filled histogram is the result of the simulation normalised to the largest values of $\theta_{flow}$. 
The open points corresponds to the difference between the data and the calculation. Down, right: same but for $E_{trans12}$. Here,
the normalisation used is the same as for $\theta_{flow}$.} 
\label{fig1}
\end{center}
\end{figure}
Anticipating the
results of the next section, we have found a good agreement with the data for the following values 
of $E^*$: 3.5 and 5. MeV/u for resp.
32 and 50 MeV/u. These values are compatible with those found in \cite{chbihi}.       
The other parameters of the simulation, that is the amount of first chance knocked-out nucleons 
and the initial multiplicities of light particles and fragments are adjusted to reproduce simultaneously the observed 
multiplicities and the center-of-mass kinetic energy distribution of the proton as shown in Fig.\ref{fig0bis}. 
The percentage of knocked-out nucleons 
increases with the beam energy as it should be since the phase space for in-medium nucleon-nucleon collisions opens 
gradually. In terms of the fraction
of the total number of nucleons, one finds: 2. and 4. $\%$ for resp. 32 and 50 MeV/u. The mean and the width of the fragment
multiplicity distribution are 6.5, 3 and 9, 3  resp. for 32 and 50 Mev/u. 
\begin{figure}[!htb]
\begin{center}
\epsfig{file=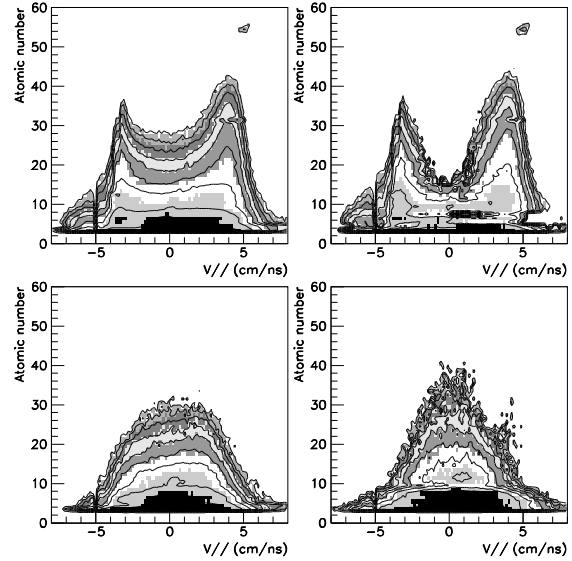,width=80mm,height=80mm}
\caption{\it 3-D plot (in log scale) of the parallel velocity vs charge of selected events
for Xe+Sn at 50 MeV/u. Up, left: experimental data. Up, right: experimental data after
'subtraction' of the simulation (see text for explanation). Down, left: experimental data for central events as defined
in the text (in this case, events with flow angles larger than 30 degrees). Down, right: same as down,left but for the simulation.} 
\label{fig2}
\end{center}
\end{figure} 
\section{Comparison with the data}
\label{dev}  
The selection of the data has been performed using first completeness criteria, requiring 
that at least 80 $\%$ of the total charge and total linear momentum (for charged particles) 
be detected. This is a necessary condition to perform an event by event analysis. 
Then, for each event, a sorting is applied by means of two variables. First, a momentum tensor analysis is developped. 
The diagonalisation of the tensor 
gives three eigen-values with which several sorting variables may be defined. Here, we use the so-called $\theta_{flow}$ 
angle. This is the angle between the main axis of the tensor and the beam axis. For more details, see \cite{adn}. 
Second, a variable using the
characteristics of the light charged particles is built: This is called $E_{trans12}$ \cite{plagnol} defined as the sum of the transverse
energies of all detected charged particles with charge lower or equal to 2.
\begin{figure}[!htb]
\begin{center}
\epsfig{file=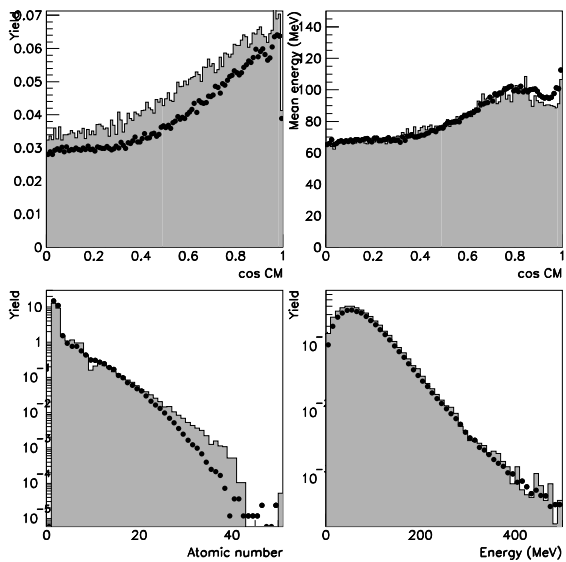,width=80mm,height=80mm}
\caption{\it Up, left: Angular center-of-mass distribution for all fragments emitted in central Xe+Sn collisions at 
50 MeV/u. Up, right: same but for CM mean kinetic energy as a function of the emission
angle. Down, left: same but for the charge
distribution. Down, right: same but for the CM kinetic energy distribution.} 
\label{fig3}
\end{center}
\end{figure} 
Fig.\ref{fig1} shows the correlation between $\theta_{flow}$ and $E_{trans12}$ for an incident energy of 50 MeV/u. 
A comparison between the
model and the experimental data shows that the model can reproduce the largest values of the $\theta_{flow}$ distribution while it misses the
low energy part. This is tentatively interpreted by assuming that the low values of the flow angle distribution are associated with
non-central collisions for which the binary character of the reaction remains until fragmentation. Indeed, in our model, this possibility is
not taken into account due to the lack of geometry of the simulation. In other words, our model in its present shape 
is only able to treat reactions corresponding to a total overlap of the two partners and is thus unable to treat deep inelatic reactions.  
{ \it From now, we consider only events for which the $\theta_{flow}$ distributions of the model and the data coincide. This corresponds 
to flow angles larger than 30 degrees for 50 MeV/u and 60 degrees for 32 MeV/u.}
\begin{figure}[!htb]
\begin{center}
\epsfig{file=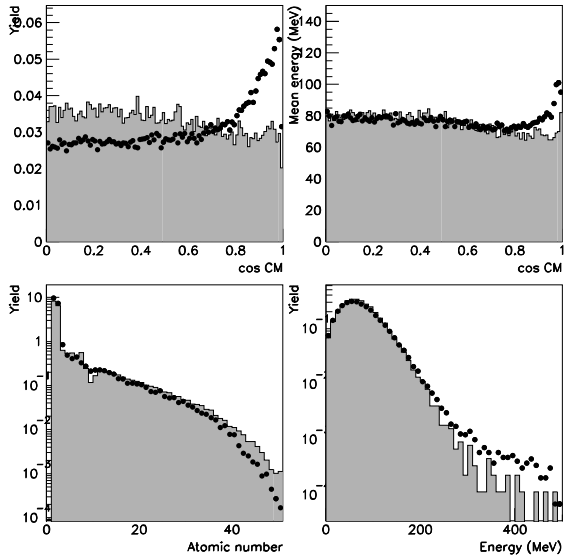,width=80mm,height=80mm}
\caption{\it Same as Fig. \ref{fig3} but Xe+Sn at 32 MeV/u.} 
\label{fig3bis}
\end{center}
\end{figure}  
All forthcoming figures have been normalised to the total number of events fulfilling this condition.   
Fig.\ref{fig2} shows the atomic number-parallel velocity plot for central Xe+Sn collisions at 50 MeV/u. The general trend of
the experimental correlation is reproduced by the simulation as shown in the lower part of the figure. Fragments are essentially emitted at
mid-rapidity. However, a closer look at the figures shows that the model underestimates the kinetic energy of the fragments around Z=20. This
point will become clearer in the following.
\begin{figure}[!htb]
\begin{center}
\epsfig{file=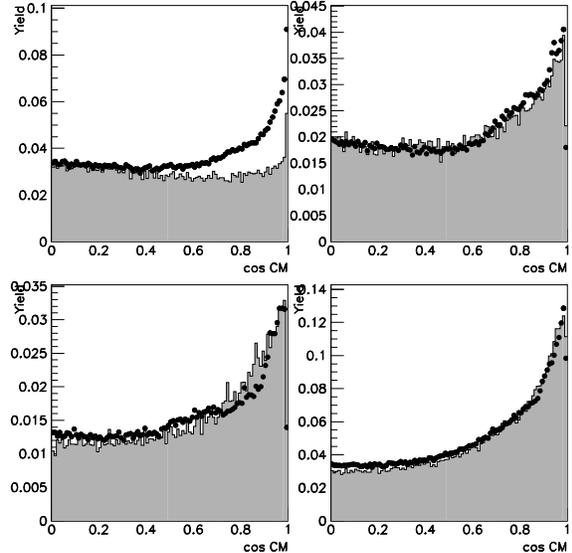,width=80mm,height=80mm}
\caption{\it  Angular centre-of-mass distributions of light particles emitted in central xe+Sn collisions at 50 MeV/u. 
Up-left: protons, up-right: deuterons, down-left: tritons, down-right: alphas} 
\label{fig5}
\end{center}
\end{figure}
A tentitative 'substraction' procedure has been applied to the data and is
displayed in the upper part of Fig. \ref{fig2}. It consists in subtracting from the experimental data 
the contribution predicted by the model and 
associated with central collisions with the normalisation discussed above.
It reveals the presence of a rather 'pure' fast component associated with the remnants of a
projectile-like and a target-like in mid-central collisions. However, for fragments with Z lower than 10, a so-called neck emission
corresponding to fragments located around mid-velocity is also
evidenced. This mechanism has already been studied for the same system but for a different selection of the data \cite{plagnol,bocage}.   
A detailed analysis of neck emission with such a technique is in progress.
We now consider some properties of the fragments in Fig. \ref{fig3} and \ref{fig3bis}. 
Due to problems with the software filter in the CM backward direction, we show only
kinematical characteristics of particles emitted in the forward part of the angular range in the CM frame. The mean number and kinetic energy
of the fragments as a function of the CM angle are rather well reproduced. The multiplicity is somehow overestimated. Here, we would like to
point out that the comparison between the model and the experiment is made without kinematical cuts: In other words, the whole event is
considered. This induces strong constraints on the model since any departure between the data and the model for some observables has
strong feed-back effects on another set of observables. 
A general 'fitting' procedure such as a back-tracing analysis should be envisaged in this context.
The kinetic energy distribution is nicely reproduced until large values showing the important role played by the internal motion of the
nucleons inside the two partners of the reaction. Same conclusions can be drawn concerning the results at 32 MeV/u (Fig.\ref{fig3bis}). 
However, here, a sizeable 
deviation in the angular distribution is observed at small angles indicating the presence of a contribution which is not accounted for in the
model. We will come back to that point later.      
\begin{figure}[!htb]
\begin{center}
\epsfig{file=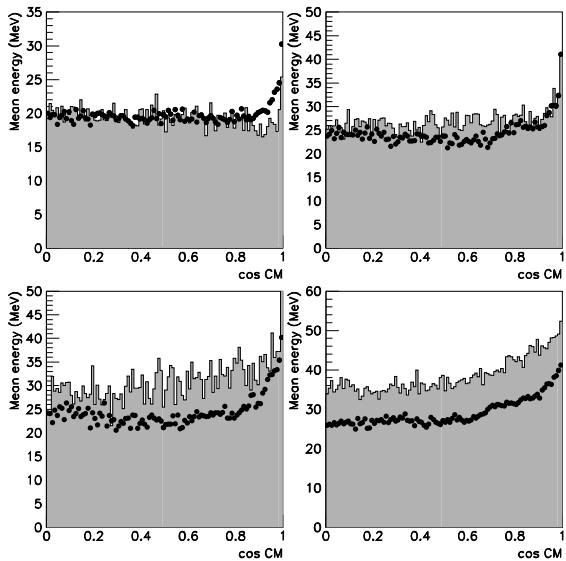,width=80mm,height=80mm}
\caption{\it Same as Fig.\ref{fig5} but for the mean kinetic energy as a function of the CM emission angle
and for Xe+Sn at 32 MeV/u.} 
\label{fig6}
\end{center}
\end{figure}
We now discuss the characteristics of light charged particles. We recall that most of them are produced by picking
randomly nucleons from the initial two Fermi spheres. However, some of them are evaporated by the excited fragments on longer time scales.
Both contributions lead to anisotropic angular distributions as shown in Fig.\ref{fig5}. We stress here that even if the final angular
distribution of the fragments may be isotropic (as is the case if one considers only events with $\theta_{flow}$ larger than 60 degrees),
those particles which have been evaporated will show an anisotropic angular distribution. Although the proton distribution is underestimated
at very forward angles, the general trend for other particles is well reproduced. 
A comparison of the mean kinetic energy as a function of the
CM emission angle (Fig.\ref{fig6}) exhibits a sizeable deviation, in particular for tritons and alphas, between the model and the data. This
means that the model is far from perfect and some aspects of the emission process for light particles is missing in our formalism. It is
however surprising that the general trends of the angular and kinetic energy distributions of light particles are rather correctly reproduced
by the model using our crude 'coalescence' algorithm.  

\begin{figure}[!htb]
\begin{center}
\epsfig{file=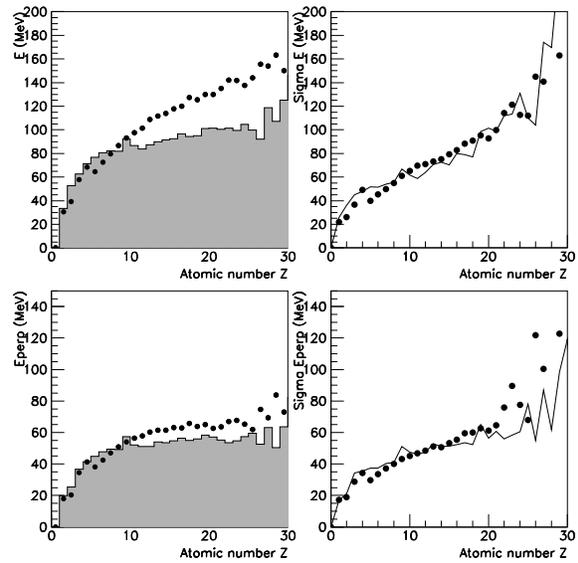,width=80mm,height=80mm}
\caption{\it Up, left: Mean kinetic energy in the center-of-mass as a function of the atomic number for fragments emitted 
in central Xe+Sn collisions at 50 MeV/u. Up, right: same but for the second moment of the kinetic energy distribution. Down, left: same as
up, left but for the transverse energies. Down, right: same as up, right but for the transverse energies.} 
\label{fig7}
\end{center}
\end{figure}

\begin{figure}[!htb]
\begin{center}
\epsfig{file=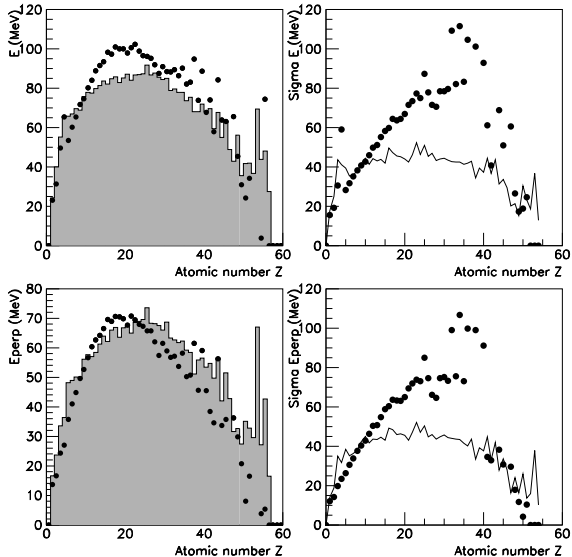,width=80mm,height=80mm}
\caption{\it Same as Fig.\ref{fig7} but for 32 MeV/u.} 
\label{fig8}
\end{center}
\end{figure}

Let us conclude this brief comparison between ELIE and the experimental data by coming back to the kinematics of the fragments.
In Fig.\ref{fig7} and \ref{fig8}, the mean and the width of the CM kinetic energy of the 
fragments as a function of their atomic number are
displayed. At 50 MeV/u, the second moment is very correctly reproduced while the mean kinetic energy is underestimated as far as 
the charge exceeds values of 15-20. This could have been anticipated looking at the results of Fig.\ref{fig2}: The memory of the entrance 
channel is
stronger for the data than for the simulation. In other words, the model (by means of the Metropolis moves described above) 
allows too much mixing between the nucleons of the target and of the
projectile as far as medium and large mass fragments are concerned. This means that an improvment of the model is needed to account
for 'geometrical' effects. In its present version, our approach can only deal with those collisions for which an almost complete overlap of
the projectile and the target is achieved. This corresponds to very small cross-sections and obviously does not cover 
the whole impact parameter
range associated with multifragmentation.
By considering only the transverse degrees of freedom (lower part of the figures), one gets rid more or less of such difficulties and the 
model then gives better results. An analysis (not shown here) of the data at larger impact parameters shows that the transverse energies of
the clusters are almost independent of the centrality of the collision except for 'controlable' final state interaction. 
Our model should be able to reproduce such trends. Work on this point is in progress.

At 32 MeV/u, a large deviation between the data and the model is observed for the widths although a rather correct agreement
is obtained for the mean values. We think the deviation is due to an additional mechanism in the data that is not considered in the model:
This is  
the remaining binary character of the collision despite the already severe selection on the impact parameter. Deep inelastic
collisions lead to a strong increase of the fluctuations in the kinetic energy distributions of the fragments, a fact that cannot be
accounted for in the present version of our model. It is worth noting that no better agreement is achieved when comparing with tranverse
quantities. This suggests that the disagreement is not only due to geometrical effects 
but to collective effects such as, for,instance, the influence of the mean field on the dynamics of the reaction.             
\section{Conclusions and perspectives}
We have studied the kinematical characteristics of nuclear products emitted in
central highly fragmented nuclear collisions at intermediate energies. 
Data have been compared with a phenomenological event generator, ELIE, in which it is assumed that 
the initial momentum distribution of the nucleons 
has no time to relax before the disassembly of the system. This distribution 
is thus used to sample the kinetic energies of the various species produced
in the reaction under the constraint of conservation laws. It is worth noting that this technique is flexible enough to test rapidly other 
(possibly thermalised) momentum distributions when the system breaks.  
As far as partitions are concerned, they are produced 'by hand' except for the charge distributions which are reproduced by a 
combinatorial random algorithm, a puzzling result which remains to be understood. 
A rather good agreement with the data has been obtained. The kinetic energy and angular 
distributions of both fragments and light charged particles are well reproduced. 

{\it In particular, in our approach, both the collective energy
and the deformation often claimed as being necessary to reproduce 
the data in the framework of thermal statistical models are 'naturally'
obtained.} 
 
We thus believe that our model is a valuable alternative to thermal 
statistical approaches 
based on shape and/or momentum equilibration in the matter at low density. 
In our approach, most light particles 
as well as fragments are produced rapidly at nearly normal density. As such, 
they keep a strong memory of the 
entrance channel. In particular, fragments emerge from the reaction highly deformed. 
Then, they decay sequentially to reach their ground state 
by evaporation on longer time scales by a rearrangement and emission of nucleons and possibly other complex particles.
The strong memory of the entrance channel and the importance of the internal motion of the nucleons inside the two partners 
of the reaction
are such that, in our picture, the origin of kinetic energy fluctuations is to a large extent non-thermal. 
We would like to point out that this picture is very similar to the one proposed in \cite{campi}.
 
Deviations between the model and the experimental data are expected as far as the incident energy increases. Work along this line is in
progress. Indeed, nuclear transparency is expected to decrease as the phase space for in-medium nucleon-nucleon collisions increases.
Therefore, the two Fermi distributions should be progressively more and more relaxed on shorter and shorter time scales 
as the incident energy increases. In the near future, we also plan to extend the model to asymetric systems and to non-central collisions.  

\end{document}